\PassOptionsToPackage{dvipsnames}{xcolor}
\documentclass[preprint,12pt, a4paper]{elsarticle}

\usepackage{amssymb}
\usepackage{hyperref}
\journal{SoftwareX}

\usepackage{xcolor}
\usepackage{float}
\restylefloat{table}
\graphicspath{{./}}
\usepackage{dirtree}

\usepackage{listings}

\DeclareFixedFont{\ttb}{T1}{txtt}{bx}{n}{9} 
\DeclareFixedFont{\ttm}{T1}{txtt}{m}{n}{9}  

\definecolor{deepblue}{HTML}{0033b3}
\definecolor{deepgreen}{HTML}{067d17}
\definecolor{commentgray}{HTML}{8c8c8c}
\definecolor{deepred}{rgb}{0.6,0,0}

\newcommand\pythonstyle{\lstset{
language=Python,
basicstyle=\ttm,
morekeywords={self},              
keywordstyle=\ttb\color{deepblue},
emph={MyClass,__init__},          
emphstyle=\color{deepred},    
stringstyle=\color{deepgreen},
commentstyle=\color{commentgray},
frame=lrtb,                         
rulecolor=\color{darkgray},
showstringspaces=false,
numbers=none,
morekeywords={as}
}}

\lstnewenvironment{python}[1][]
{
\pythonstyle
\lstset{#1}
}
{}
\newcommand\pythoninline[1]{{\pythonstyle\lstinline!#1!}}

\begin{document}
\renewcommand{\labelenumii}{\arabic{enumi}.\arabic{enumii}}

\begin{frontmatter} 

\title{Traffic Weaver: semi-synthetic time-varying traffic generator based on averaged time series}

\author[pwr,chalmers]{Piotr Lechowicz}\corref{cor1}
\ead{piotr.lechowicz@pwr.edu.pl}
\cortext[cor1]{corresponding author}
\address[pwr]{Department of Systems and Computer Networks, Wroc\l{}aw University of Science and Technology, Wroc\l{}aw, Poland}
\address[chalmers]{Department of Electrical Engineering, Chalmers University of Technology, Gothenburg, Sweden}

\author[pwr]{Aleksandra Knapi\'nska}
\author[pwr]{Adam W\l{}odarczyk}
\author[pwr]{Krzysztof Walkowiak}

\begin{abstract}
{Traffic Weaver is a Python package developed to generate a semi-synthetic signal (time series) with finer granularity, based on averaged time series, in a manner that, upon averaging, closely matches the original signal provided. The key components utilized to recreate the signal encompass oversampling with a given strategy, stretching to match the integral of the original time series, smoothing, repeating, applying trend, and adding noise. The primary motivation behind Traffic Weaver is to furnish semi-synthetic time-varying traffic in telecommunication networks, facilitating the development and validation of traffic prediction models, as well as aiding in the deployment of network optimization algorithms tailored for time-varying traffic.}

\end{abstract}


\end{frontmatter}



\begin{table}[!h]
\begin{tabular}{|l|p{6.5cm}|p{6.5cm}|}
\hline
\textbf{Nr.} & \textbf{Code metadata description} & \textbf{Please fill in this column} \\
\hline
C1 & Current code version & 1.3.5 \\
\hline
C2 & Permanent link to code/repository used for this code version &  \url{https://github.com/w4k2/traffic-weaver} \newline \url{https://pypi.org/project/traffic-weaver/}\\
\hline
C3  & Permanent link to Reproducible Capsule & For example: \\
\hline
C4 & Legal Code License   & MIT \\
\hline
C5 & Code versioning system used & git \\
\hline
C6 & Software code languages, tools, and services used & Python \\
\hline
C7 & Compilation requirements, operating environments \& dependencies &  Python $\ge$ 3.9 \\
\hline
C8 & If available Link to developer documentation/manual & \url{http://w4k2.github.io/traffic-weaver/} \\
\hline
C9 & Support email for questions & piotr.lechowicz@pwr.edu.pl \\
\hline
\end{tabular}
\caption{Code metadata (mandatory)}
\label{codeMetadata} 
\end{table}



\section{Motivation and significance}



In telecommunication networks, such as backbone optical networks, many small end-to-end transmissions between individual users and devices  combine into time-varying traffic, representing aggregated traffic over time. Thus, daily and weekly patterns can be observed in network traffic due to increased user activity in certain periods. Driven by the paradigm of self-driving and self-healing networks, traffic prediction, and anomaly detection gained significant research community attention in recent years.
However, the community faces the problem of lacking real data, allowing for thorough experiments. Network operators are often constrained by legal aspects and cannot share the details of traffic generated by their customers. In turn, many researchers can have access either to small exemplary data or to averaged data without sufficient quality. 
To this end, the community relies on artificially generated data with various distributions and patterns based on their domain knowledge (e.g., \cite{parsonson2022traffic,valkanis2021traffic,wlodarczyk2020algorithm,petale2023prodigy}). However, predicting and detecting changes in real data can bring significantly more challenges than artificially generated ones. Additionally, extensive experiments performed on a large pool of appropriately diverse datasets are necessary for the development and thorough evaluation of the designed algorithms \cite{hoffmann2019benchmarking}.

The purpose of Traffic Weaver is to generate new data based on an already available sample of data, i.e., to create semi-synthetic data when the size of real data is either insufficient or the time points at which the data were measured are too rare. In particular, the software has been used in scientific research to create semi-synthetic time-varying traffic to develop and evaluate traffic prediction models \cite{knapinska2022prediction,ulanowicz2023combining} and multi-layer network optimization algorithms using generated time-varying connection requests (intents) \cite{knapinska2023advantages,knapinska2023performance}. Semi-synthetic data allowed a thorough evaluation of the developed algorithms in real-world settings and various desired characteristics. 

The aim of Traffic Weaver is to read averaged time series and to create a~semi-synthetic signal with finer granularity that, after averaging, matches the original signal provided.
The following tools are applied to recreate the signal: oversampling with a given strategy, stretching to match the integral of the original time series, smoothing, repeating, applying trend, and adding noise. Software users may provide exemplary data for the investigated problem or use one of the available small datasets to create semi-synthetic time-varying traffic by applying various trends and noise profiles. The increased input data size allows for a more thorough investigation of the problem. Moreover, the ability to create datasets with specific characteristics enables detailed testing of the developed algorithms in various conditions \cite{stapor2021design}.

\section{Software description}


\begin{figure}[!ht]
    \centering
    \includegraphics[width=.8\textwidth]{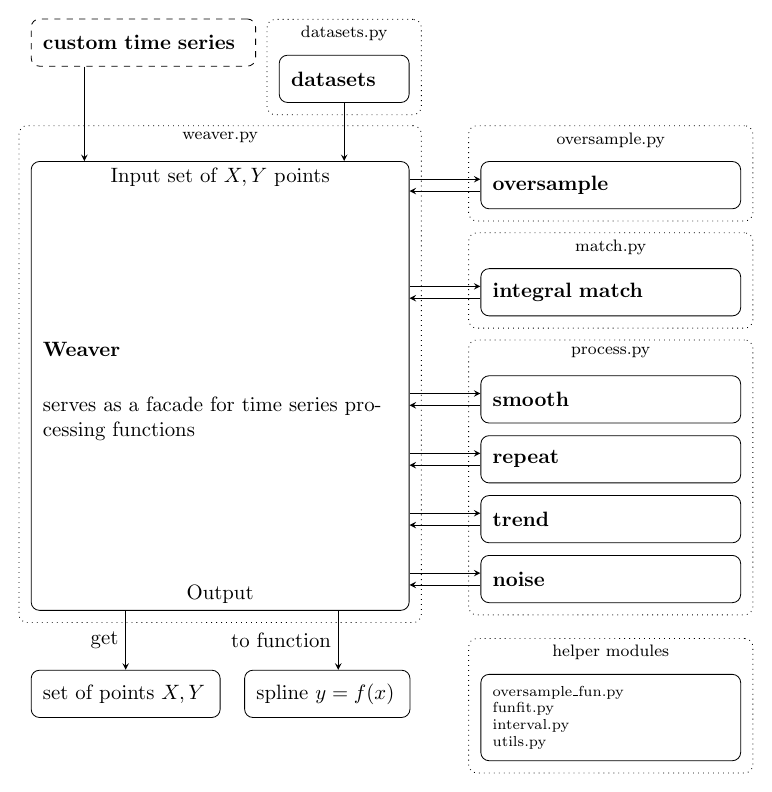}
    \caption{Software architecture.}
    \label{fig:software-architecture}
\end{figure}

\subsection{Software architecture}

Fig.~\ref{fig:software-architecture} presents an overview of the software architecture. \textit{Weaver}, located in \textit{weaver.py} module, wraps supplied signal (time series) data and provides an interface for processing functionalities. Time series can be either specified by the user or obtained from embedded example datasets.
Individual functionalities provided by the \textit{Weaver} are delegated to other modules, e.g., oversampling functionality is located in the \textit{oversample.py} module. However, it is possible to use individual functionalities from the corresponding modules regardless of wrapping time series into \textit{Weaver}. \textit{Weaver} allows retrieving the processed data either as sampled points or as a continuous spline function.



\subsection{Software functionalities}

This section describes the main functionalities provided by the Traffic Weaver. In the below description, the term \textit{interval} refers to the distance between two sampled points in the input time series. The aim of the Weaver is to create an output time series with multiple points inserted in each interval.

\begin{itemize}
\item Class \textit{Weaver(x, y)}

    \textit{Weaver} is an interface for recreating signal. It takes as an input time series provided as two lists containing values of independent and dependent variables. It delegates processing to other modules and allows to retrieve the recreated signal either as lists of values of independent and dependent variables or as a spline, using \textit{get()} and \textit{to\_function()} methods, respectively.

\item Oversampling 

    \sloppy
    Oversampling is a recreation of a signal with finer sampling granularity based on the supplied strategy. The number of created points between each interval (pair of points in the original time series) is provided as a parameter. The strategy determines how the created time series transits between points, i.e., how the new points are located. The software provides several strategies, namely, \textit{ExpAdaptiveOversample()}, \textit{ExpFixedOversample()}, \textit{LinearAdaptiveOversample()}, \textit{LinearFixedOversample()}, \textit{PiecewiseConstantOversample()}, \textit{CubicSplineOversample()}. E.g., \textit{ExpAdaptiveOversample()} creates an adaptive transition window for each interval by combining linear and exponential functions. The size of the window is inversely proportional to the change of the function value on both edges of the interval, i.e., if the function value has a higher change on the right side than on the left side of the interval, the right side transition window is smaller than the left one.

    The \textit{Weaver} class provides the \textit{oversample(n, oversample\_class, **kwargs)} method that delegates the execution to the oversample module and takes as an input number of samples \textit{n} in each interval after oversampling, oversample strategy \textit{oversample\_class} inheriting \textit{AbstractOversample()} class, and a dictionary of parameters passed to the selected strategy.

\item Integral matching 

    It aims to reshape the time series to match its integral to the integral of the reference piecewise constant function over the same domain (the original time series). It does that by stretching the signal in intervals such that the integral in the interval of the current time series is equal to the integral of the same interval in the reference function. Points in each interval are transformed inversely proportionally to the exponential value of distance from the interval center.

    The \textit{Weaver} class provides the \textit{integral\_match(**kwargs)} method that delegates the execution to the match module and takes as an input a dictionary of parameters passed to the matching function. The time series currently stored in the Weaver is matched with a reference to the originally passed function to the class.

\item Smoothing

    It smooths a function using smoothing splines.

    The \textit{Weaver} class provides the \textit{smooth(s)} method to delegate the execution to the smoothing function and takes \textit{s} as an argument. The argument \textit{s} is a smoothing condition that controls the tradeoff between closeness and smoothness of the fit. Larger \textit{s} means more smoothing, while smaller values of \textit{s} indicate less smoothing. If \textit{s} is None, its ``good'' value is calculated based on the number of samples and standard deviation.

\item Repeating

    It repeats time series a given number of times, resulting in a long term time series containing periodic, e.g., daily or weakly, patterns.

    The \textit{Weaver} class provides the \textit{repeat(n)} method to repeat the time series. \textit{n} is an argument passed to the function, defining how many times to repeat the time series.

\item Trending

    It applies a trend to the time series according to the specified function. It allows adding a long-term trend to the time series, e.g., constant dependent variable increase over time.

    The \textit{Weaver} class provides the \textit{trend(trend\_func)} method to apply a trend to the processed time series. The argument \textit{trend\_function} is a callable that shifts the value for the dependent variable based on the value of the independent variable normalized to a $(0, 1)$ range. The callable takes one argument – the normalized value of the independent variable – and has to return the shift value for the dependent variable.

\item Noising

    It applies a constant or changing over time Gaussian noise to the time series, expressed as signal to noise ratio.

    The \textit{Weaver} class provides the \textit{noise(snr, **kwargs)} method to apply noise to the signal. The argument \textit{snr} defines the signal-to-noise ratio of a function either as a scalar value or as a list of changing values over time whose size matches the size of the independent variable. \textit{**kwargs} is a set of parameters passed to the noising function, allowing, e.g., to express the noise as a normal distribution standard deviation instead of the signal-to-noise ratio.

\item Datasets

    The \textit{Datasets} module provides example datasets based on the Sandvine report \cite{Sandvine2021}. In more detail, \cite{Sandvine2021} includes information about daily traffic patterns of various network-based applications, e.g., TikTok, YouTube, Zoom, etc., averaged over multiple large networks. The report presents the data as bar plots of traffic averaged in each hour of the day. The \textit{Datasets} module includes nineteen datasets containing information about these shapes denoted as lists, which can be used as a base for generating the semi-synthetic traffic.

\end{itemize}
  

\section{Illustrative examples}



\begin{figure}[!ht]
    \centering
    \includegraphics[width=\textwidth]{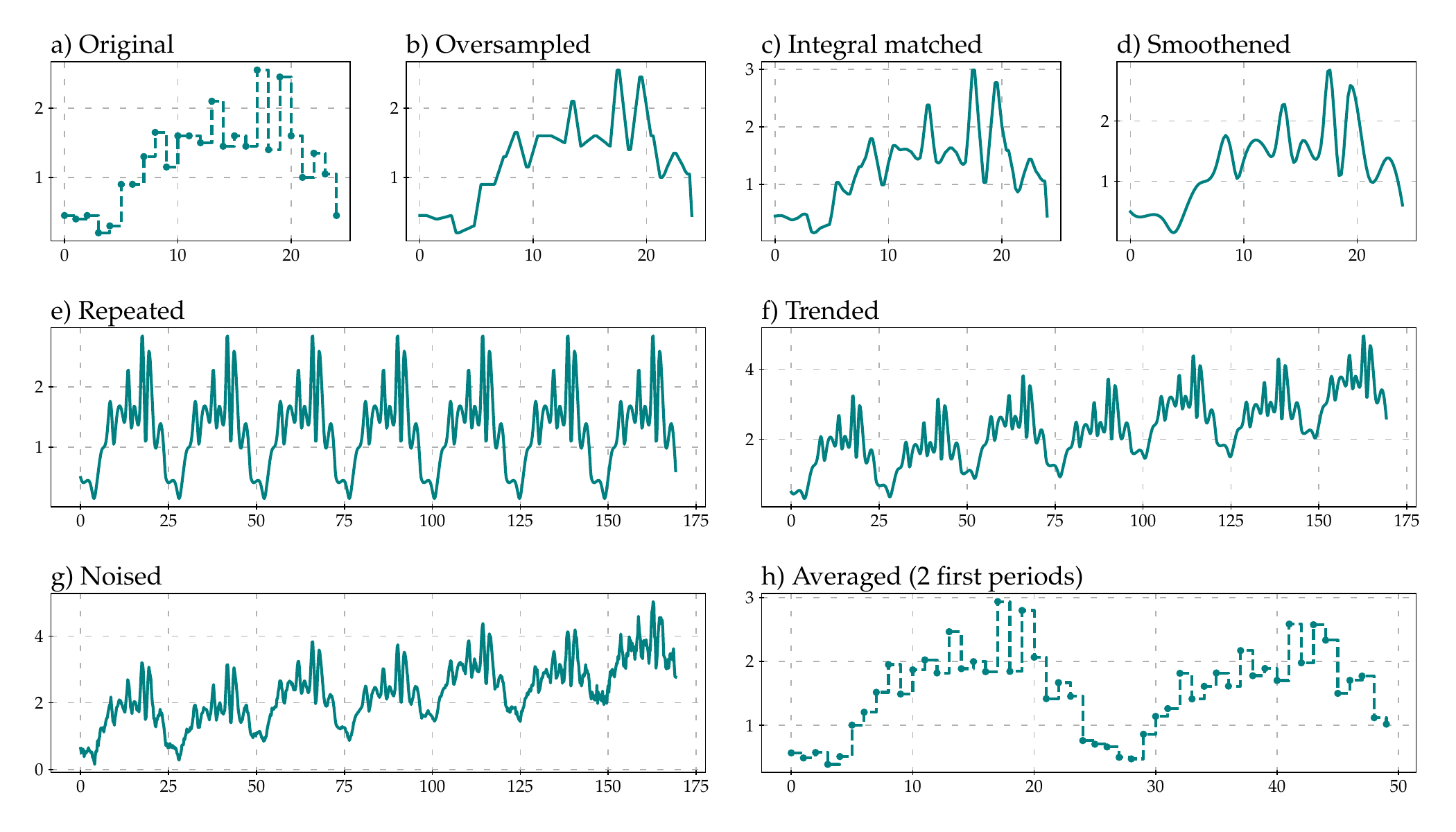}
    \caption{Illustrative example.}
    \label{fig:signal-processing}
\end{figure}

Fig.~\ref{fig:signal-processing} shows a general usage example. Based on the provided original averaged time series (a), the signal is $n$-times oversampled with a predefined strategy (b). Next, it is stretched to match the integral of the input time series function (c). Further, it is smoothed with a spline function (d). In order to create weekly semi-synthetic data, the signal is repeated seven times (e), applying a long-term trend consisting of sinusoidal and linear functions (f). Finally, the noise is introduced to the signal, starting from small values and increasing over time (g). To validate the correctness of the applied processing, (h) presents the averaged two periods of the created signal, showing that they closely match the original signal (except the applied trend).

\subsection{Minimal processing example}

\textit{Traffic Weaver} is an open source Python module released under MIT license and versioned in the public \textit{Python Package Index} (PyPI) repository. It can be installed using \textit{pip} package manager.

\begin{lstlisting}[language=sh]
pip install traffic-weaver
\end{lstlisting}

Traffic Weaver import is done with the standard import command.

\begin{python}
import traffic_weaver
\end{python}

To load one of the exemplary datasets, use one of functions provided in the \textit{datasets} module. 

\begin{python}
# load example dataset with average measurements over 1 hour
x, y = traffic_weaver.datasets.load_tiktok()
\end{python}

The \textit{traffic\_weaver} module provides the \textit{Weaver} class that serves as an API to other processing capabilities. The \textit{Weaver(x, y)} constructor takes the time series independent and dependent variables as arguments, denoted as \textit{x} and \textit{y}, respectively.

\begin{python}
# create Weaver instance
wv = traffic_weaver.Weaver(x, y)
\end{python}

Further signal processing is applied through Weaver methods. Most of the methods return an instance to the Weaver itself, allowing for chaining the processing commands.

\begin{python}
# process it creating samples every minute
wv.oversample(60).integral_match().smooth(1.0).noise(snr=30)
\end{python}

\begin{figure}[!ht]
    \centering
    \includegraphics[width=\textwidth]{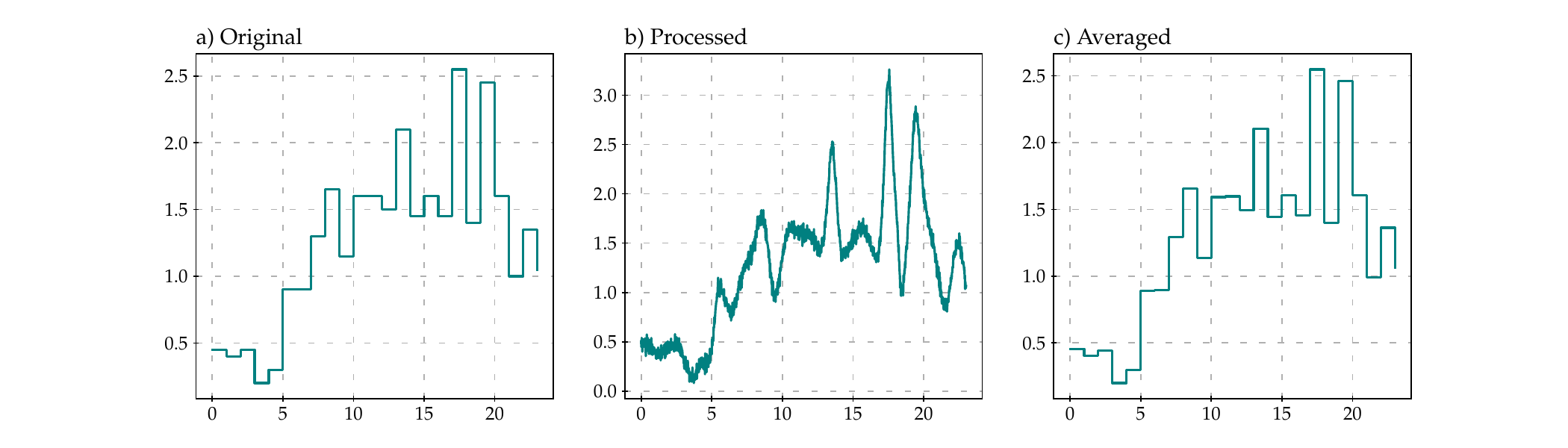}
    \caption{Minimal processing example.}
    \label{fig:minimal-processing-example}
\end{figure}

{\sloppy
To obtain the created new time series, call either \textit{Weaver's} \textit{get()} or \textit{to\_func\-tion()} methods. Next, visualize time series with matplotlib - the original data, created semi-synthetic traffic, and averaged semi-synthetic traffic to verify that the integral of semi-synthetic traffic does not differ much from the one in original signal. The result of the below listing is presented in Fig.~\ref{fig:minimal-processing-example}.}

\begin{python}
import matplotlib.pyplot as plt

# plot original signal
fig, axes = plt.subplots(nrows=1, ncols=3, figsize=(14, 4))
axes[0].plot(*wv.get_original(), drawstyle="steps-post")

# plot modified signal
axes[1].plot(*wv.get())

# plot averaged signal
x, y = traffic_weaver.process.average(*wv.get(), 60)
axes[2].plot(x, y, drawstyle="steps-post")

axes[0].set_title("a) Original", loc="left")
axes[1].set_title("b) Processed", loc="left")
axes[2].set_title("c) Averaged", loc="left")
plt.show()
\end{python}

\section{Impact}

The networking community lacks a public data repository for research purposes and the development of new optimization methods based on network traffic. Existing analyses of real traffic data (e.g., \cite{garcia2012characterization,jurkiewicz2021flow,goscien2021modeling}), collected by the authors over long periods, usually stop at the data characterization stage and are not further used in the networking research nor are they easily accessible. 
Traffic Weaver closes this gap, allowing easy access to data and enabling thorough evaluation of developed algorithms. Using various options provided in Traffic Weaver, the created methods can be tested in diverse traffic conditions representing actual traffic patterns. In turn, the package allows a fair and versatile algorithm development, evaluation, and comparison with the existing solutions. It also helps in gaining insights into the operation of various methods in specific traffic conditions considering parameters such as noise levels, trends, and traffic type. These parameters are impossible to steer using the available sparse raw data. 
Moreover, Traffic Weaver is implemented in Python, which is the primary programming language used for the development of machine learning methods \cite{nguyen2019machine}. 

\section{Conclusions}
This article presents Traffic Weaver – a semi-synthetic time-varying traffic generator. The software creates new datasets based on either existing examples of real data or user-specified data and enables adding desired characteristics. Through a variety of processing methods, including oversampling, integral matching, smoothing, repeating, trending, and noising methods, the package allows a thorough evaluation of created optimization and prediction methods based on the network traffic. The available example datasets provide a versatile entry for networking research.

\section*{Acknowledgements}
\label{}
This work was supported by the National Science Center, Poland under Grant 2019/35/B/ST7/04272.

\bibliographystyle{elsarticle-num} 
\bibliography{bibliography}

\end{document}